\documentclass[final,1p,times]{elsarticle}

\usepackage{graphicx}
\usepackage{subfigure}
\usepackage[rightcaption]{sidecap}

\usepackage{amssymb}

\journal{Nuclear Physics A}

\begin{document}

\begin{frontmatter}


\ead{schumacher@cmu.edu}

\title{Hyperon Photo- and Electro- Production Experiments at CLAS}

\author{Reinhard Schumacher \\
(for the CLAS Collaboration)}
\address{Department of Physics, Carnegie Mellon University, Pittsburgh, PA 15213, USA}

\begin{abstract}
Developments in strangeness photo- and electro- production off the
proton, as investigated using the CLAS system in Hall B at Jefferson
Lab, are discussed in this paper. By measuring sufficient spin
observables one can decompose the reaction mechanism into elementary
amplitudes.  We discuss progress toward this end in recent data from
CLAS, including cross sections and spin observables.  We next discuss
new results on the mass distribution of the $\Lambda(1405)$, which shows
signs of being a composite meson-baryon object of mixed isospin.  The
work on other hyperons such as the $\Xi$ resonances will be mentioned,
and future prospects outlined.

\end{abstract}

\begin{keyword}
hyperons \sep photoproduction \sep electroproduction \sep CLAS \sep spin observables
\sep $\Lambda(1405)$ \sep baryon resonances
\PACS 14.20.Jn  \sep 25.20.Lj \sep 14.20.Gk \sep 13.30.-a



\end{keyword}

\end{frontmatter}


\section{Introduction}
\label{intro}
The CLAS~\cite{Mecking:2003zu} detector system in Hall B at the Thomas Jefferson
National Accelerator Laboratory is a multi-purpose, multi-particle
spectrometer for photo- and electro-excitation of mesons and baryons.
The scientific goal is to understand the spectrum of non-strange baryon
excitations (``$N^*$ physics''), including the spin structure, transition
form factors, and decay modes of these states.  To this end, polarized
and tagged real photon beams with energy up to 5.7 GeV, and electron
scattering for $Q^2$ up to about 4.5 GeV/c$^2$ have been used in
conjunction with proton, deuteron, and nuclear targets.  The targets
have been polarized longitudinally and (in the future) transversely to
the incident beams.

As is well known, the cross section for elementary production of
hyperons and strange mesons is of the order of a few microbarns in the
c.m. energy range from threshold at 1.6 GeV to about 2.5 GeV.  This is
a percent-level contribution to the overall production rate dominated
by single and multiple pion final states.  In the search for
resonances predicted in quark models~\cite{Capstick:1998uh} but unseen
in experiments~\cite{PDG}, the small contributions by strange-particle
final states could be decisive.  First, in the mass range of
``missing'' resonances above 1.6 GeV, the production of hyperons in
reactions such as $\gamma p \rightarrow K^+ \Lambda$ is comparatively
easy to detect and analyze.  The dominant multi-pion final states are
more complicated to handle theoretically and experimentally.  Second,
the self-analyzing aspect of hyperon decays allows for the
straightforward measurement of polarization observables involving the
recoiling baryon, much more so than for non-strange final states where
polarization measurements require a secondary scattering.  In fact,
there is a prospect at CLAS to make ``complete'' measurements of
enough spin observables in the hyperon production channels that the
entire amplitude structure can be determined.  This has not been done
yet for any elementary pseudo-scalar meson photo-production process.

On the other hand, the smallness of the hyperon electromagnetic
production cross sections relative to the non-strange channels means
that channel coupling effects are significant. To understand the $N^*$
spectrum in photo- and electro-excitation requires modeling of these
effects to enforce unitarity and to gain realistic estimates of the
couplings for particular channels.  Such efforts are underway, see for
examples Refs.~\cite{tmart} and ~\cite{bg}.

Beside production and decay of $N^*$ and $\Delta^*$ resonances into
the ground state $\Lambda$ and $\Sigma^0$, CLAS can study the
production of hyperon resonances such as $\Sigma(1385)$,
$\Lambda(1405)$, $\Lambda(1520)$, etc., and also the $\Xi$ resonances.
The interest in these states is that some may turn out not to be
elementary quark model states, but rather be dynamically generated
objects made through the resonant re-interaction of a ground state
pseudoscalar meson and an octet baryon, as discussed for example in
Ref.~\cite{osetHYP}.

Most of the results shown during this conference were preliminary, but
should be published in archival journals within the next months.  Some
of the key new results are included here.

\section{Cross Sections and Spin Observables}
\label{cs}
Elementary photoproduction of any pseudo-scalar meson, including
kaon-hyperon (``$KY$'') production, can be described by four complex
amplitudes that define the reaction completely.  Bilinear combinations
of these amplitudes lead to 16 observables that require specification
of some combination of the polarization of the incident photon, the
nucleon target, and the recoiling baryon~\cite{barker}.  The
differential cross section is unique in that it depends only on the
sum of the amplitude magnitudes squared.  
The others involve various algebraic 
combinations of the amplitudes and of interference terms among them.
The
problem of finding a minimal set of measurements to totally separate
the amplitude components at any given energy was solved~\cite{tabakin}
to show that eight of sixteen observables must be measured, but not
any eight will do.  Beside the cross section $d\sigma/d\Omega$, the
three single spin observables of beam asymmetry ($\Sigma$), target
asymmetry ($T$), and recoil polarization ($P$) are needed, as are four
combinations of double-spin observables involving the three available
spins.  To date, the goal of a ``complete'' set of measurements on any
elementary production reaction channel has not been achieved.
However, it is the ambition of the CLAS program to reach this goal for
the $\gamma p\rightarrow K^+\Lambda$ and $\gamma p\rightarrow
K^+\Sigma^0$ channels within the next few years. There is similar hope
that enough measurements can be made for the neutron-target channels
as well, so as to pin down the isospin dependence of the reactions.

One must distinguish between determination of the four reaction
amplitudes and understanding the reaction mechanism of strangeness
photoproduction.  Once the amplitudes are measured well enough, in a
sense the task of the experimentalists is finished since no other
information can be gleaned from any other measurements.  The task of
the model-builders is not done, of course.  For example the $N^*$
content of the intermediate states in the $s$-channel leading to the
$KY$ final state remains to be parsed from the amplitudes.

\begin{figure}[htpb]
\centering 
\subfigure[]{\includegraphics[width=3.5 cm,angle=90.0]{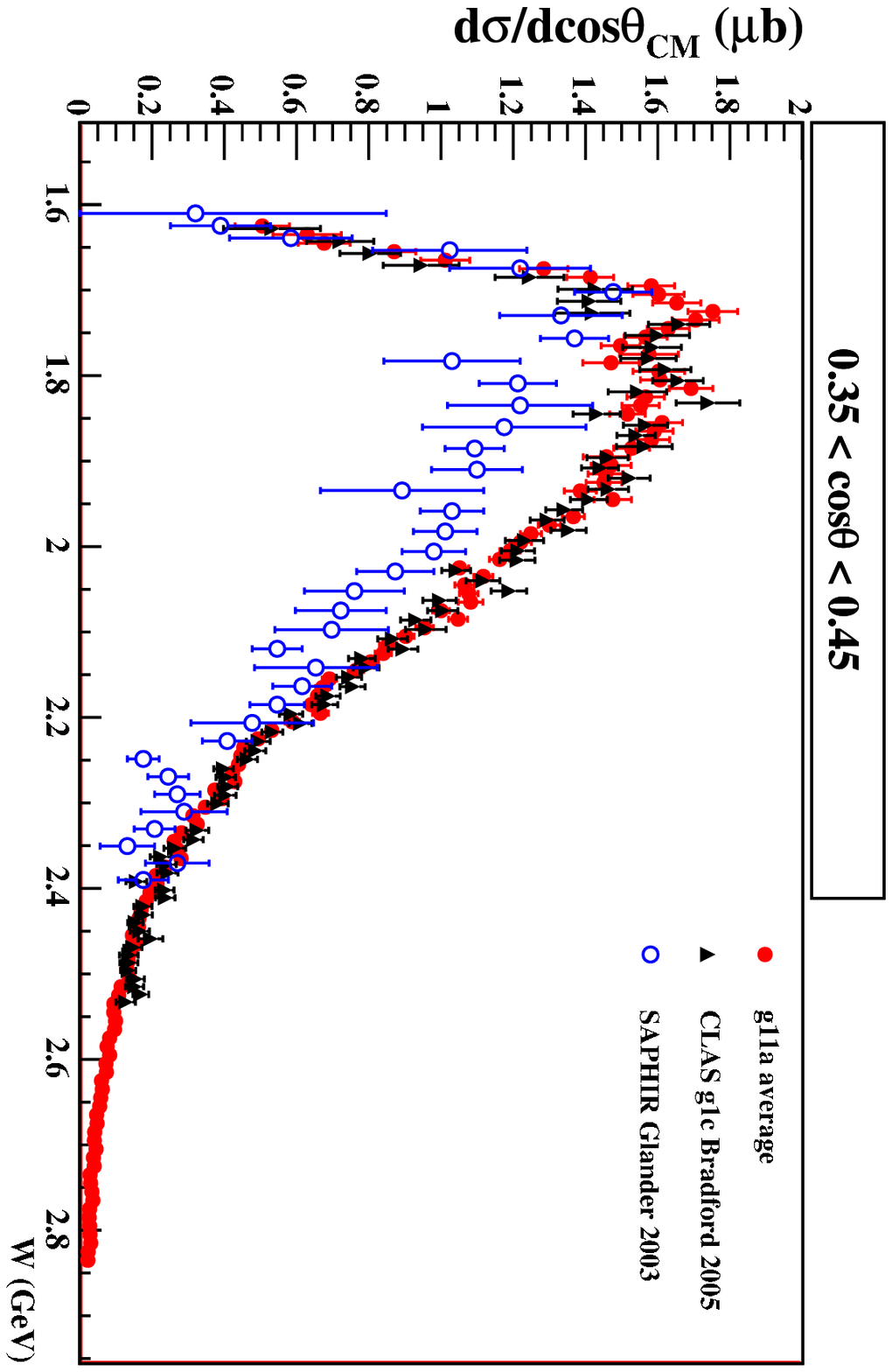}}
\subfigure[]{\includegraphics[width=3.5 cm,angle=90.0]{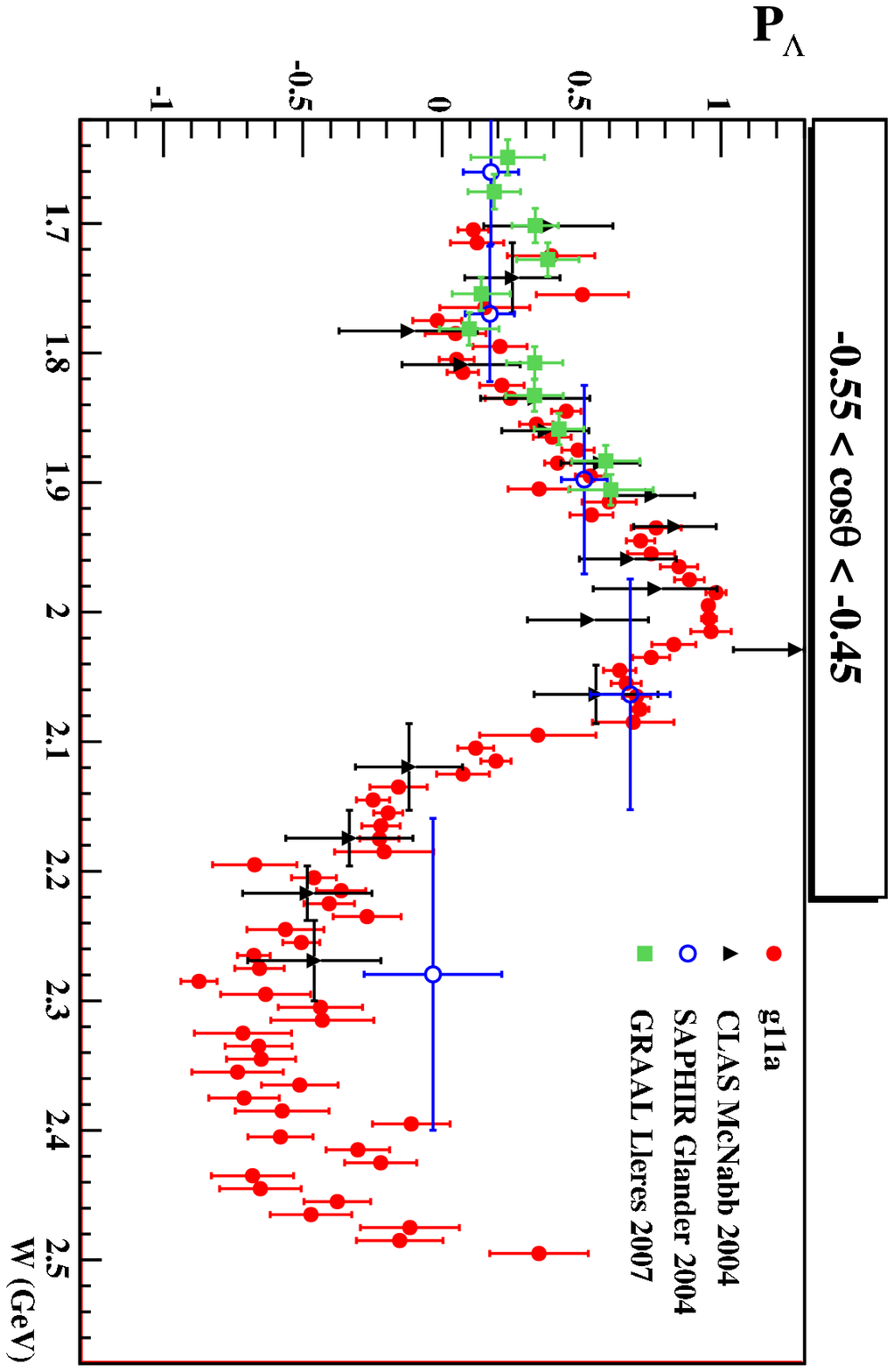}}
\caption{(Color online) (a) Comparison of published differential cross
sections for $\gamma p \rightarrow K^+ \Lambda$ from
CLAS~\cite{Bradford:2005pt} (blue triangles),
SAPHIR~\cite{Glander:2003uf} (open blue circles), and new (2009)
results from CLAS~\cite{mccracken} (closed red circles). The
c.m. production angle range of the $K^+$ is given in the figure title.
(b) Comparison of published $\Lambda$ recoil polarization for $\gamma
p \rightarrow K^+ \Lambda$ from CLAS~\cite{Bradford:2005pt},
SAPHIR~\cite{Glander:2003uf}, GRAAL~\cite{Lleres:2007tx}, and new
results from CLAS~\cite{mccracken}.  }
\label{fig:mccracken}
\end{figure}

CLAS has published photoproduction cross sections for the reactions
$\gamma p \rightarrow K^+ \Lambda$ and $\gamma p \rightarrow K^+
\Sigma^0$ over a broad angular range and at energies up to 2.54
GeV~\cite{Bradford:2005pt}.  There has been
controversy~\cite{Bydzovsky:2006wy} about a discrepancy of these
measurements with those from SAPHIR at Bonn~\cite{Glander:2003uf}.
Near $W=1.9$ GeV, both experiments reported a bump-like feature in the
cross section that is clearly the signature of an underlying baryonic
structure. It is either a resonance, with quantum numbers that are
open to debate (see for example Refs.~\cite{tmart, Nikonov:2007br}),
or perhaps it is the signature of a $\overline{K}KN$
dynamically-generated resonance~\cite{MartinezTorres:2009cw,osetHYP}.
There is good agreement between CLAS and LEPS~\cite{leps_sumihama} at
forward angles.  New measurements at CLAS have been performed using
the so-called ``g11'' data set, obtained for pentaquark searches, but
which has provided copious data for other reaction channels.
Figure~\ref{fig:mccracken}(a) shows a sample of these data, showing,
in one kaon production angle bin, the improved statistics and wider
energy coverage compared to previous results.  The agreement with the
published CLAS results is excellent, and the discrepancy with SAPHIR
is evidently resolved in the favor of CLAS.  Obviously the new results
are not total independent measurements, since they use the same CLAS
detector, but the data set, the experimental trigger, the analysis
chain, the Monte Carlo methods, and the people doing the analysis were
all different.

CLAS also published results for the out-of-plane hyperon recoil
polarization $P$ ~\cite{McNabb:2003nf} that were in good agreement
with those produced by SAPHIR~\cite{Glander:2003uf} and by
GRAAL~\cite{Lleres:2007tx}.  As shown in Fig.~\ref{fig:mccracken}(b),
the new results from CLAS are in excellent agreement with the earlier
work, and have higher precision and broader energy coverage.  It is
remarkable how some ``fine structure'' is newly visible in the $P$
results.  There is no definite interpretation yet of these structures,
pending an on-going partial wave analysis project at Carnegie
Mellon~\cite{mccracken}; the data alone will be published separately.

Using coherent bremsstrahlung from a diamond radiator, linearly
polarized photons were produced between 1.15 and 2.05 GeV.  CLAS has
measured the beam asymmetry ($\Sigma$) in the
$K^+\Lambda$~\cite{paterson} channel (not shown), which is the
azimuthal sine-like variation of the production intensity.  Results
are in agreement with published data from GRAAL~\cite{Lleres:2007tx}
and LEPS~\cite{leps_zegers}, but cover a broader energy and angular
range.  Also using linearly polarized photons, CLAS has measured the
beam-recoil observables for transfer of the transverse polarization of
the photon to the recoiling hyperon~\cite{paterson}.  This was done by
detecting the parity-violating weak decay of the hyperons, measured in
each bin of kaon production angle and incident beam energy.  The
observables are called $O_x$ and $O_z$, corresponding to polarization
transfer in the scattering plane.  One can choose these axes to align
either with the beam axis or the kaon-hyperon axis in the c.m. frame,
since the polarization transforms like a three-vector.  The
preliminary results show that near threshold the $\Lambda$ hyperon is
nearly fully polarized ``sideways'', as one might guess naively, but
the hyperon ``picks'' to be polarized pointing toward the beam-axis
side of the reaction plane.  No models predict this phenomenon.

The incoming photons can be circularly polarized via bremsstrahlung
from longitudinally polarized electrons; in CLAS this electron beam
polarization can be around 80\%.  This gives access to the beam-recoil
observables representing the transfer of the photon helicity to the
produced hyperon in the reaction plane.  These observables are called
$C_x$ and $C_z$, again with freedom in picking the axes aligned with
either the beam or the final-state particles.  CLAS has measured these
observables~\cite{Bradford:2006ba}, and found that for a wide range of
kinematics the photon helicity is transferred very much along the beam
direction, corresponding to $C_z\simeq1$.  But even more unexpected
was that the three components of the recoiling hyperon polarization,
$C_x$, $C_z$, and $P$, when produced with circularly polarized
photons, summed in magnitude to essentially unity.  That is, $R=\sqrt{
C_x^2+C_z^2+P^2} \simeq 1$ over the whole of the measured angle and
energy range.  This has been experimentally confirmed by recent data
from GRAAL~\cite{Lleres:2008em}.  This phenomenon was described in a
semi-classical model~\cite{Schumacher:2008xw} at the previous HYP
conference.  However, since then several groups have incorporated this
phenomenology into various effective Lagrangian models.

The accumulation of recent CLAS, LEPS, GRAAL, and SAPHIR results for
$KY$ photoproduction has been used in several model-building studies.
The ``resonance'' at 1.9 GeV was initially interpreted by Bennhold and
Mart~\cite{bennholdmart} as a $D_{13}$ state in the Kaon-MAID model,
as inspired by the search for missing resonances as specified by
Capstick and Roberts~\cite{Capstick:1998uh}.  In a recent updating of
this model, Mart finds~\cite{tmart} more consistency with assigning
this structure the quantum numbers $P_{13}$.  In another instance, the
Bonn-Gatchina group has shown~\cite{bg,Nikonov:2007br} that, as part
of their elaborate multi-channel unitary baryon excitation picture,
the $C_x$ and $C_z$ results strongly favor the excitation of a
$P_{13}$ resonance with a pole at 1915 MeV.  This would qualify as
finding a ``missing'' quark model state.  It is inconsistent with at
least one quark-diquark model of baryon structure~\cite{santopinto}.
On the other hand, there is a suggestion that this large peak in the
$K\Lambda$ cross section is not a standard baryon resonance at all,
but rather a $\overline{K}KN$ bound
state~\cite{MartinezTorres:2009cw}.

In elementary electroproduction of strangeness via the reactions
$\vec{e}+p\rightarrow e^\prime+K^+ + Y$, the unseparated
electroproduction cross sections shows a bump-like structure in
$\Lambda$ channel near 1.9 GeV, just as in
photoproduction~\cite{Ambrozewicz:2006zj}.  CLAS published results for
the transfer of longitudinal polarization of the virtual photon to the
produced hyperon~\cite{Carman:2002se}.  Newer work shows that the
phenomenon of large spin transfer along the (virtual) beam direction
remains valid over a quite broad kinematic range, for $W$ up to at
least 2.31 GeV and $Q^2$ out to 2.41 (GeV/c)$^2$~\cite{Carman:2009fi}.
The helicity asymmetry in electron scattering, $\sigma_{LT}^\prime$,
sometimes called the ``fifth structure function'' in electron
scattering from an unpolarized target, has also been recently
published~\cite{Nasseripour:2008fz}, and it covers the region of the
bump at 1.9 GeV.  The conclusion of that work was to disfavor the
$P_{11}$ interpretation of the structure.

\section{The $\Lambda(1405)$}
\label{1405}

\begin{SCfigure}[][htpb]
\centering 
\includegraphics[width=7.0 cm,angle=0.0]{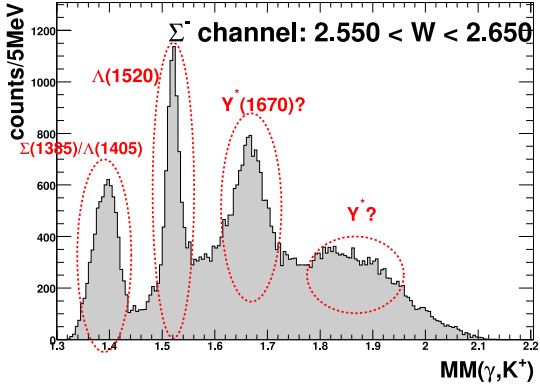}
\caption{(Color online) Sample CLAS data showing the production of
  several hyperon states via photoproduction off the proton.  The
  reaction in this case is $\gamma p \to K^+ Y \to K^+ \Sigma^- \pi^+
  \to K^+ \pi^- (n) \pi^+$, with given c.m. energy W.  Shown is the
  missing mass off the $K^+$ integrated over all production angles.  }
\label{fig:rawlambda}
\end{SCfigure}

The $\Lambda(1405)$ hyperon is a well-known state, but its internal
structure has been problematic for five decades.  While it may contain
a 3-quark component, it is widely discussed as a candidate for a
$\overline{K}N$ molecular state, or as a dynamically generated resonance made
via meson-baryon rescattering~\cite{Nacher:1998mi,Borasoy:2005ie}.
Figure~\ref{fig:rawlambda} is the raw hyperon spectrum seen with CLAS
in $\gamma + p \rightarrow K^+ + Y$ when $Y \rightarrow \Sigma^- \pi^+
\rightarrow \pi^- (n) \pi^+$, where the parenthetical neutron is
reconstructed by missing mass.  Clear signals for the low-lying
hyperons are seen on top of a background.  The combination of
$\Lambda(1405)$ and $\Sigma(1385)$ is visible, as well as a large
$Y^*$ signal near 1670 MeV and possible bumps near 1850 MeV.  CLAS can map
out this excitation with a resolution of $\sigma\simeq6$ MeV.  The
background is due to particle mis-identification, and in other decay
channels the physics background from $K^*$ production.

\begin{figure}[htpb]
\centering 
\subfigure[]{\includegraphics[width=6.5 cm,angle=0.0]{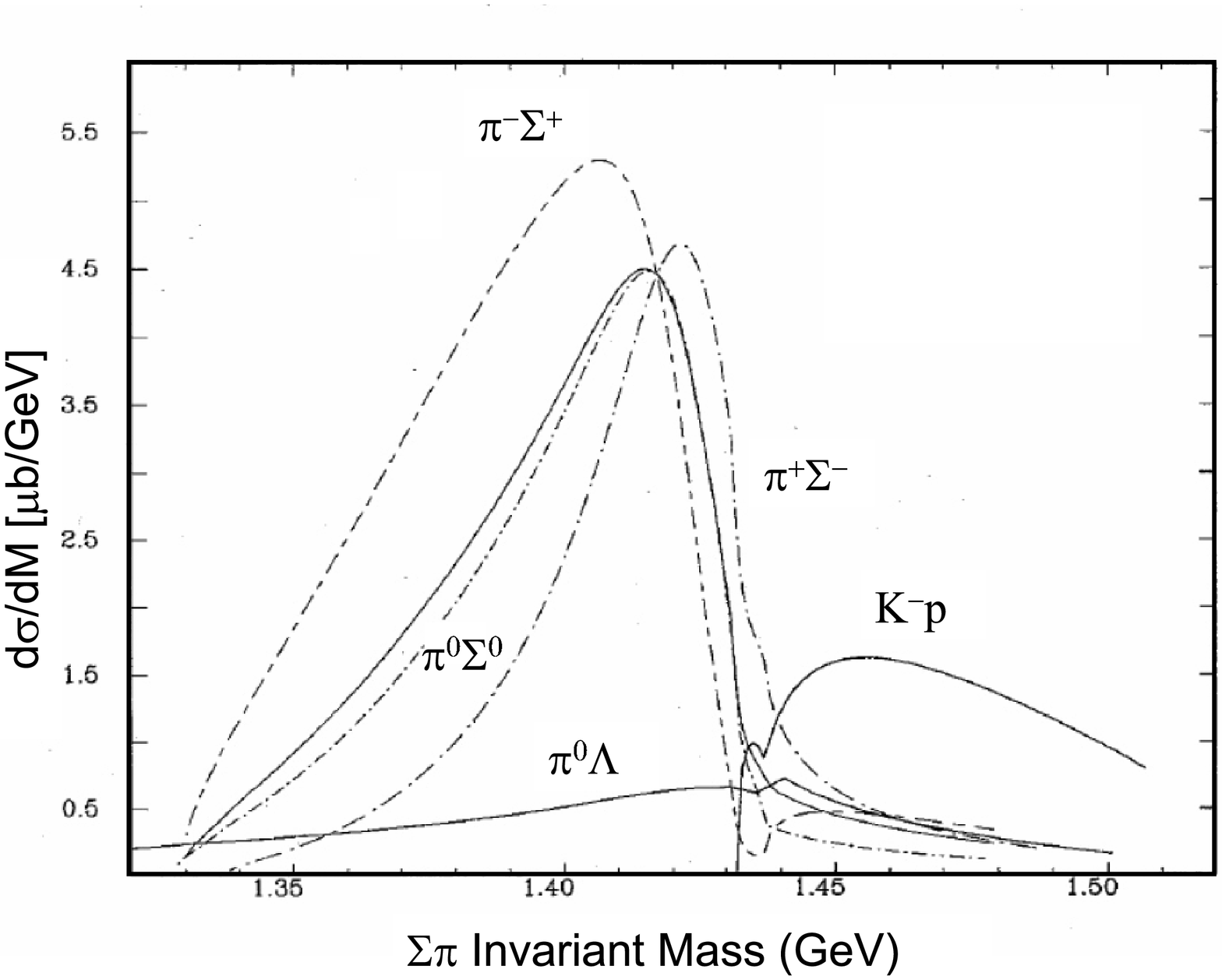}}
\subfigure[]{\includegraphics[width=6.5 cm,angle=0.0]{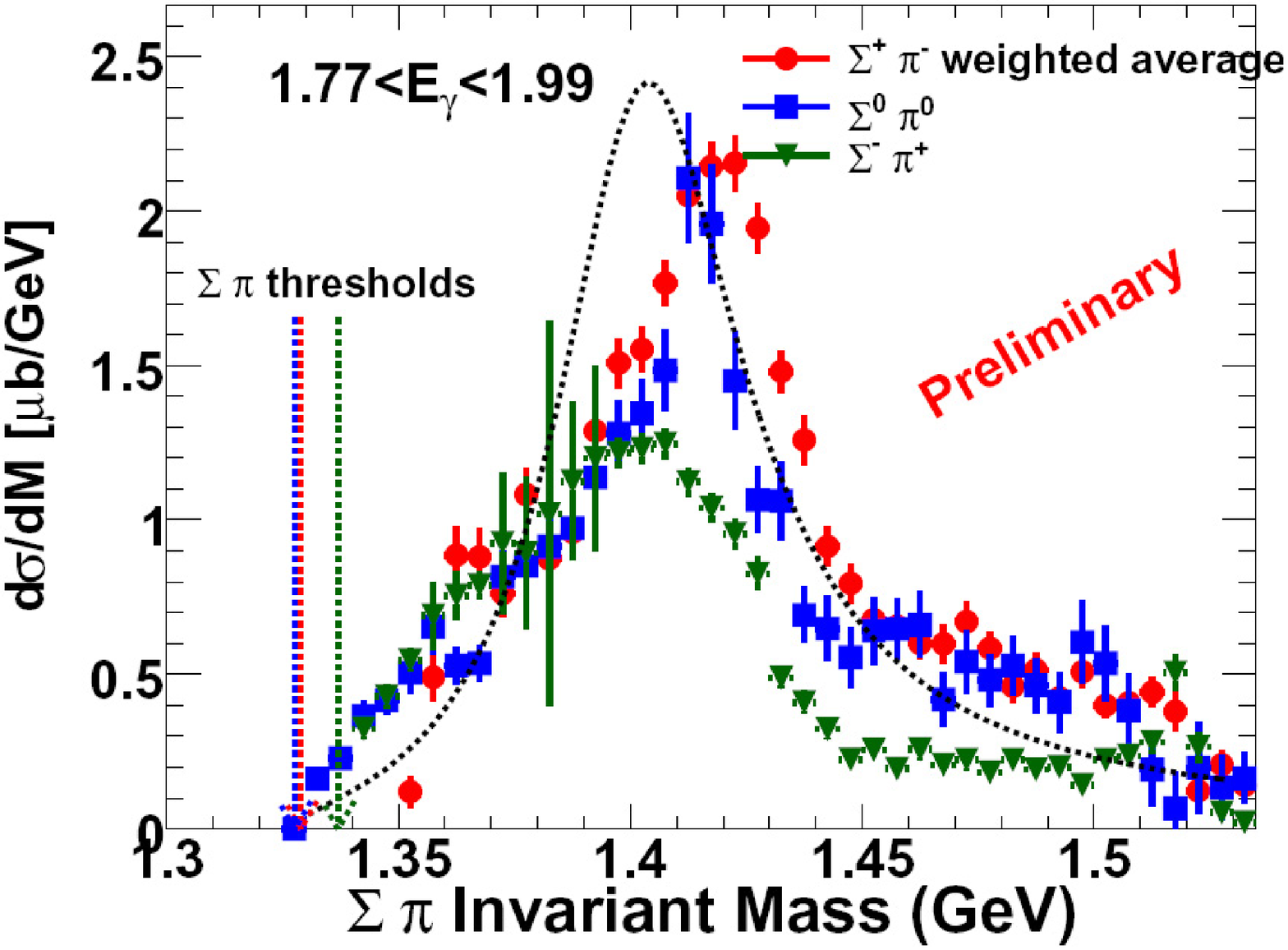}}

\caption{(Color online) (a) Predicted ~\cite{Nacher:1998mi} line
  shapes of the $\Lambda(1405)$ hyperon in several decay channels for
  $E_\gamma=1.7$ GeV.  In this model the $\Lambda(1405)$ is generated
  dynamically by resonances in isospin 0 and 1.  (b) Preliminary mass
  distributions of the $\Lambda(1405)$ in each of its three different
  decay channels in an energy bin near threshold. The different decay
  modes indeed show a difference in shape and strength. The dotted
  line is a relativistic Breit- Wigner line shape drawn using the
  nominal $\Lambda(1405)$ mass and width as given in the
  RPP~\cite{PDG}.}

\label{fig:lambda1405}
\end{figure}

If the $\Lambda(1405)$ were a conventional 3-quark state of
well-defined isospin, the mass distribution or ``line shape'' would be
the same for charge-symmetric decays $\Lambda(1405) \to \Sigma^+\pi^-$
and $\Lambda(1405) \to \Sigma^-\pi^+$, as well as for $\Lambda(1405)
\to \Sigma^0\pi^0$.  That is, one would expect a basically
Breit-Wigner line shape, with distortions expected due to the opening
of the $\overline{K}N$ decay channel about 27 MeV above the peak of
the structure and small isospin-breaking mass differences.  However,
there is a prediction by the Valencia group~\cite{Nacher:1998mi} that
the $\Lambda(1405)$ is a dynamically generated resonance with
contributions from two isospin amplitudes. The interference of $I=0$
and $I=1$ amplitudes will cause the $\Sigma\pi$ mass distributions to
differ in a specific way, as shown in Fig.~\ref{fig:lambda1405}(a).
In other models, two poles in the $I=0$ channel are discussed, one
mostly coupling to $KN$ and the other to
$\Sigma\pi$~\cite{Borasoy:2005ie}; these would also distort the BW
line shape.  Preliminary photoproduction evidence for the plausibility
of these surmises came from the LEPS group at
SPring-8~\cite{Ahn:2003mv}. However, that experimental result was
statistically weak.  Another recent result from COSY at Julich,
looking at the $\Lambda(1405)$ in $pp$ collisions through the
$\Sigma^0\pi^0$ decay mode, suggested that there is no apparent
difference among the decay modes to different charge final
states~\cite{Zychor:2007gf}.  Again, this result was weak in
statistics.  What is needed is a high-statistics sample of this
hyperon decaying to several charge states, and this is what CLAS can
provide.  If mass distribution distortions are confirmed, this would
be an example of a well-known baryon resonance shown to have a
structure fundamentally outside of the usual valence quark model and
its extensions.

The $\Lambda(1405)$ photoproduction differential cross section and
line shape was recently measured again by the LEPS
collaboration~\cite{Niiyama:2008rt} at very forward kaon angles ($0.8
< \cos \theta_{K^{+}}^{CM} < 1.0$) and energies of $1.5 < E_{\gamma} <
2.0$ and $2.0 < E_{\gamma} < 2.4$ GeV. Their extraction of the
$\Lambda(1405)$ differential cross section assumed a theoretical
$\Lambda(1405)$ line shape, and their results had poor statistics. The
CLAS detector and g11 data set allowed reconstruction of the line
shape of the $\Lambda(1405)$ directly from the data, without relying
on theoretical models, with sufficient statistics to
acceptance-correct the line shape in each bin.

The mass distribution and differential cross section of the $\Lambda(1405)$
in all of its decay channels have been extracted.  A preliminary
sample result is shown in Fig.~\ref{fig:lambda1405}(b). This is the
$\Lambda(1405)$ line shape for each of its decay modes $\frac{d
\sigma}{d M}$ in units of $\mu b\mathrm{GeV}^{-1}$ for $1.95 < W <
2.05$ GeV. All distributions have been acceptance corrected for detector
losses and summed over all production angles.  Separation of various
components of the line shape was achieved by fitting the total
spectrum with Monte Carlo data~\cite{moriya}, and therewith
subtracting off contributions from $K^*$ production and $\Sigma(1385)$
production (scaled from its decay to the $\Lambda\pi^0$ final
state). Results in nine $W$ bins that CLAS covered in the g11 run
period are pending.

In the figure, notice that the $\Sigma^{+} \pi^{-}$ mass distribution
starts well above the threshold, and has the most narrow peak at the
highest mass. The $\Sigma^{-} \pi^{+}$ distribution has the broadest
structure, while the $\Sigma^{0} \pi^{0}$ distribution comes in between
the charged $\Sigma \pi$ line shapes. This is in contrast to the
theory of ~\cite{Nacher:1998mi}, where it is the $\Sigma^{-} \pi^{+}$
line shape that is predicted to be the one with the highest mass at
its peak and the most narrow structure.  Nevertheless, it is notable
how the theory captures at least the qualitative aspects of the
isospin-induced modifications of the production line shape.  Detailed
systematic studies are in progress to finalize these spectra.  Since
the effects are quite strong in the data, the next step will be to
estimate the contributions from the $I=0$ and $I=1$ components of the
reaction.

The parity of the $\Lambda(1405)$ has not been measured before and the
spin needs confirmation, though no one would be very surprised if the
answer were $1/2^-$, as predicted by the quark model.  The CLAS data
allows a direct measurement of the spin and parity, leading to the
expected result.  Some additional details of these measurements are
given in Ref.~\cite{moriya}.

\section{Other Hyperon Resonances}
\label{others}

CLAS can measure photoproduction of other $S=-1$ excited hyperons.
Some results awaiting archival publication include data for the
$\Lambda(1520)$ and $\Sigma(1385)$.  The main motivation for such
studies at this point is to fill in the very scanty database of
information for these reactions.  Electroproduction of the
$\Lambda(1520)$ has been reported~\cite{Barrow:2001ds}, showing that
in the $t$-channel the exchanged mesons are a mixture of spin 0 and
spin 1, unlike in photoproduction, where the exchange is dominantly
spin 1.  Photoproduction of the $K^*(892)$ in the reaction $\gamma p
\rightarrow K^{*0}\Sigma^+$ has been reported~\cite{Hleiqawi:2007ad}
and compared with some success to a simple effective Lagrangian model.

The $S=-2$ hyperons have received attention since they can be detected
in CLAS through reactions of the type $\gamma p\rightarrow K^+ K^+
\Xi^-$ or $\gamma p\rightarrow K^+ K^+ \pi^- \Xi^0$.  In quark models,
the $\Xi$ resonances should be as numerous as the $N^*$ states.  They
are small in production cross section, but have the good feature that
they are expected to be fairly narrow; for example, the first excited
state, the $\Xi(1530)$, has been seen in photoproduction, and its
roughly 9 MeV width is to be contrasted with the $\simeq 100$ MeV
width of any $N^*$ state.  CLAS has reported~\cite{guo}
photoproduction of $\Xi^-(1321)$, $\Xi^0(1315)$, and $\Xi^{*-}(1530)$.
Fair agreement was found between the measured ground state cross
section and an effective Lagrangian model~\cite{nakayama} based on the
notion of $t$-channel excitation of high-mass $S=-1$ states decaying
to the ground state $S=-2$ $\Xi$ hyperon.  Finding ``new'' high mass
$S=-2$ states is the goal of on-going analysis of a CLAS data set with
a 5.7 GeV endpoint, as well as eventual new experiments in the 12 GeV
era at Jefferson Lab.  Small but statistically insignificant hints of
higher mass $\Xi$'s in the published data~\cite{guo} are tantalizing.

\section{Future Prospects}
\label{future}
There are three categories of measurements still in the ``pipeline''
at CLAS that will lead to additional hyperon-related results.  First,
with a deuteron target one can access the elementary production cross
sections off the neutron, and a large data set (``g13'') was
accumulated in 2006 with circularly polarized photons up to 2.6 GeV,
and in 2007 with linearly polarized photons up to 2.3 GeV.  It is
expected that processes such as $\gamma n (p) \rightarrow K^0 Y^0 (p)$
or $\rightarrow K^+ \Sigma^- (p)$ will be accessible, where $(p)$
designates the spectator proton.  Information from the other isospin
channels in elementary photoproduction is naturally of interest to
gain the complete understanding of these reactions.  Quasi-free
reactions on the proton can also be observed.

Second, the CLAS frozen spin butanol target, FROST, was operated
successfully with longitudinal polarization in 2007, and processing of
this data is well along.  $K\Lambda$ spin observables such as $E$ and
$T$ are being determined.  Using the holding field of a ``saddle
coil'', it is planned to run with transversely polarized butanol in
the spring and summer of 2010, so that the dream of a ``complete'' set
of spin observables for $KY$ photoproduction will be achievable.

Finally, the remaining goal at CLAS is to have a polarized neutron
target to measure the corresponding spin observables.  A target system
called ``HD-Ice'' is under development which pumps molecular HD
material such as to selectively polarize just the deuterium or just
the hydrogen.  This system is planned to run in CLAS late in 2010 and
the first part of 2011.

\section{Conclusion}
\label{conclusion}

This brief overview of recent strangeness physics results from
Jefferson Lab's CLAS facility should illustrate the variety of issues
being addressed.  Improved precision of several hyperon photo- and
electro-production observables is making the modeling of the
$K\Lambda$ and $K\Sigma^0$ channels more exact and conclusive.  This
is relevant to the search for non-strange quark-model baryon
resonances missing in the experimental record.  The dream of measuring
enough observables to construct the complete set of amplitude-level
components of these channels is within reach, using existing and
upcoming data using polarized targets (FROST, HD-Ice), polarized
photon beams (both circular and linear), and detecting the recoil
hyperon polarization.

The $\Lambda(1405)$ was shown, in preliminary results, to have
significant isospin-mixing effects in its decay to $\Sigma\pi$, which
is relevant to the study of the pole structure on the $S=-1$
meson-baryon system near threshold.

The lightest $\Xi$ hyperons have been detected and measured in
photoproduction, and work is in progress to look for other $S=-2$
states with higher mass using data sets obtained with beam energies
approaching 6 GeV.

Between 2009 and the end of the 6 GeV era at CLAS in late 2011, more
data sets will be acquired using transverse proton target polarization
and also polarized neutron (deuterium) targets.  Thus, it is likely
that CLAS will be able to present additional results on the production
and decay of $KY$ systems into the middle of the coming decade.

\end{document}